\RequirePackage{fix-cm}
\documentclass[smallextended]{svjour3}       
\smartqed  
\usepackage{graphicx}
\usepackage{lineno}
\usepackage{xcolor}

\modulolinenumbers[2]

\def\gtorder{\mathrel{\raise.3ex\hbox{$>$}\mkern-14mu
 \lower0.6ex\hbox{$\sim$}}}
\def\ltorder{\mathrel{\raise.3ex\hbox{$<$}\mkern-14mu
 \lower0.6ex\hbox{$\sim$}}}

\begin{document}

\title{Searching for three-nucleon short-range correlations}


\titlerunning{3N-SRCs: Status and future plans}    

\author{Nadia Fomin \and John Arrington \and Shujie Li}

\institute{N. Fomin \at
              University of Tennessee, Knoxville, Knoxville, TN 37996, USA\\
           \and
           J. Arrington \and S. Li \at
              Lawrence Berkeley National Laboratory, Berkeley, CA 94720, USA \\
}

\date{Received: date / Accepted: date}

\maketitle

\begin{abstract} ABSTRACT

Electron scattering measurements from high-momentum nucleons in nuclei at SLAC and Jefferson Lab (JLab) have shown that these nucleons are generally associated with two-nucleon short-range correlations (2N-SRCs). These SRCs are formed when two nucleons in the nucleus interact at short distance via the strong tensor attraction or repulsive core of the NN potential. A series of measurements at JLab have mapped out the A dependence and isospin dependence of 2N-SRCs, and have begun to map out their momentum structure. However, we do not yet know if 3N-SRCs, similar high-momentum configurations of three nucleons, play an important role in nuclei. We summarize here previous attempts to isolate 3N-SRCs, go over the limitations of these previous attempts, and discuss the present and near-term prospects for searching for 3N-SRCs, mapping out their A dependence in nuclei, and constraining their isospin and momentum structure.

\end{abstract}

\section{Introduction} \label{sec:intro}

The starting point for experimental studies of short-range correlations (SRCs) were inclusive measurements~\cite{Frankfurt:1993sp} that showed scaling in scattering cross section ratios from high-momentum nucleons in nuclei from the deuteron to gold. This confirmed the predictions~\cite{Frankfurt:1981mk,Frankfurt:1988nt} of the naive SRC model, in which this universal behavior arises from the hard, short-distance components of the NN interaction. The short-range tensor attraction or repulsive core interaction generates nucleons in nuclei with momenta $k$ well above the Fermi momentum, causing the $k>k_{Fermi}$ contributions to have a universal structure, driven by the nature of the NN interaction.

Later measurements, mainly at Jefferson Lab, mapped out the A dependence of the SRC contributions~\cite{Fomin:2011ng,CLAS:2003eih,egiyan06,CLAS:2019vsb}, and demonstrated that these two-nucleon SRCs (2N-SRCs) were dominated by neutron-proton (np) pairs, and that pp- and nn-SRCs had much smaller contributions~\cite{Subedi:2008zz,LabHallA:2014wqo,CLAS:2018xvc,JeffersonLabHallA:2007lly}.  Ref.~\cite{Arrington:2022sov} provides a broad overview of the experimental status of these measurements, and discusses future experimental plans to go beyond our current understanding of SRCs in nuclei.

In this work we focus on recent measurements and future plans to go beyond 2N-SRCs by searching for and studying the structure of three-nucleon short-range correlations (3N-SRCs). These represent configurations where three nucleons have large relative momenta but modest total momentum, generated by either a pair of hard two-nucleon interactions or by a hard three-nucleon interaction~\cite{Day2023}.

The first step is to compare inclusive scattering from a heavy nucleus to a 3-body nucleus.  This allows to look for a 3-body contribution to the momentum distribution that may become dominant after the 2N-SRC contributions are sufficiently small. This would indicate the ability to isolate contributions from 3N-SRCs, as was done previously for 2N-SRCs.  If this can be successfully accomplished, the next step would be to try and understand the isospin and momentum structure of these configurations.
\section{What we learned from 2N-SRC searches}
\begin{figure}[htb] 
\centering
  \includegraphics[width=0.98\textwidth]{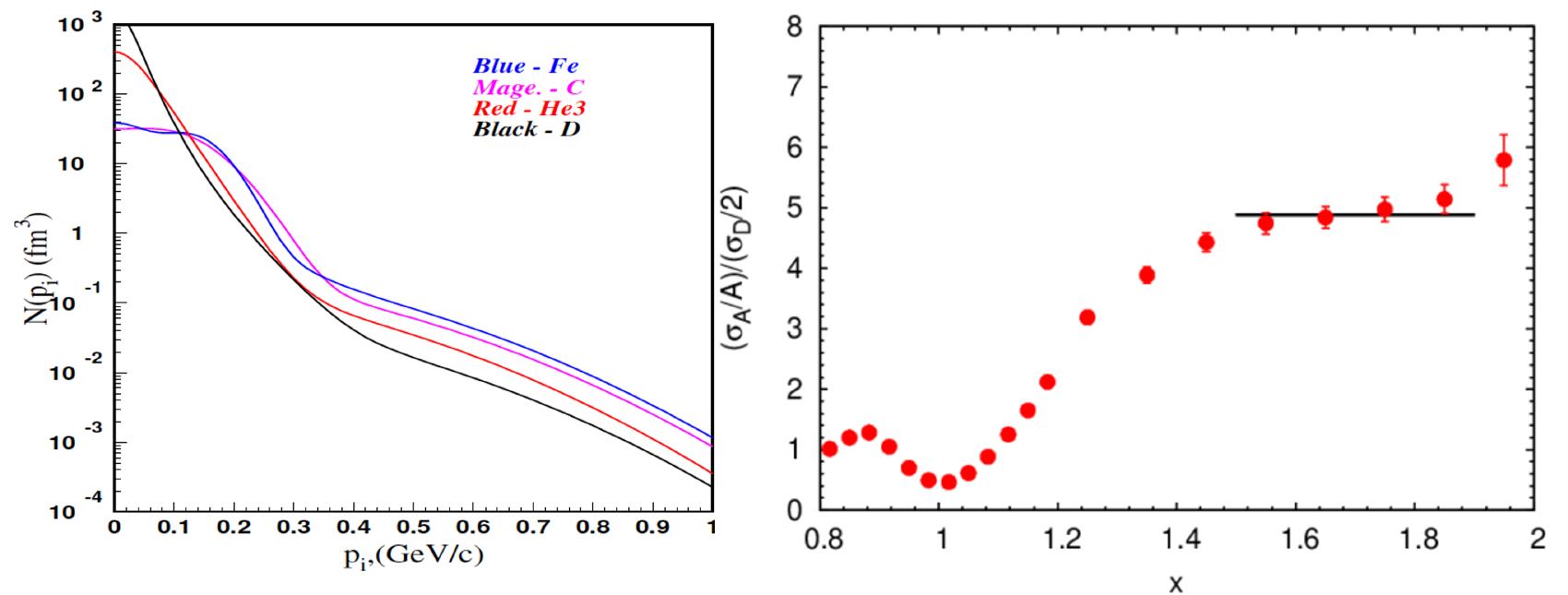}
  \caption{LEFT: Calculated momentum distributions for several nuclei, adapted from~\cite{CiofidegliAtti:1995qe}. RIGHT: Cross section ratio for Carbon, data from~\cite{Fomin:2011ng}.}
  \label{fig:2nplat}
  \end{figure}  
The high-momentum tail contribution to the inclusive cross section can be written as~\cite{Frankfurt:1993sp}:
\begin{equation}
    \sigma(x,Q^2)=\sigma_{MF}+\sum_{j=2}^{A} A\frac{1}{j}a_j(A)\sigma _j(x, Q^2),
\end{equation}
where $\sigma_{MF}$ is the mean field single particle contribution, $a_j$ are the probabilities of finding correlations of $j$ nucleons, and $\sigma_j$ is the cross section for scattering from a correlation of $j$ nucleons.  To isolate the contribution from 2N correlations in inclusive cross section ratios, one must reach a region where the mean field strength falls off enough for the contribution of NN tensor force to dominate.  Typically, this is expected to start at nucleon momenta of about 300~MeV/c~(Fig.~\ref{fig:2nplat}, left), with the 2N-SRC contribution beginning to dominate. For a given value of $Q^2$, an onset in $x$ can be calculated~\cite{Fomin:2008iq}.  Past that onset, the $\sigma_A /\sigma _D$ cross section ratio is expected to exhibit a scaling plateau with a value of $a_2$. This is illustrated in Fig.~\ref{fig:2nplat}, where the plot on the left-hand side shows calculated momentum distributions for a number of nuclei, with the high-momentum tail shape being similar across all, as it is driven by the same, short-range NN tensor interaction. 

For $A=2$, the kinematic limit is $x\sim$2.  However, for larger nuclei, the kinematic limit continues up to $x\sim A$, but the cross section continues to falls off very quickly (exponentially). Measures of cross section ratios of $A>2$ nuclei in the $x>2$ region could yield evidence of 3N SRCs, but their isolation and data interpretation is more challenging.
\section{Challenges in 3N-SRC Searches}

\begin{figure}[htb] 
\centering
  \includegraphics[width=0.49\textwidth]{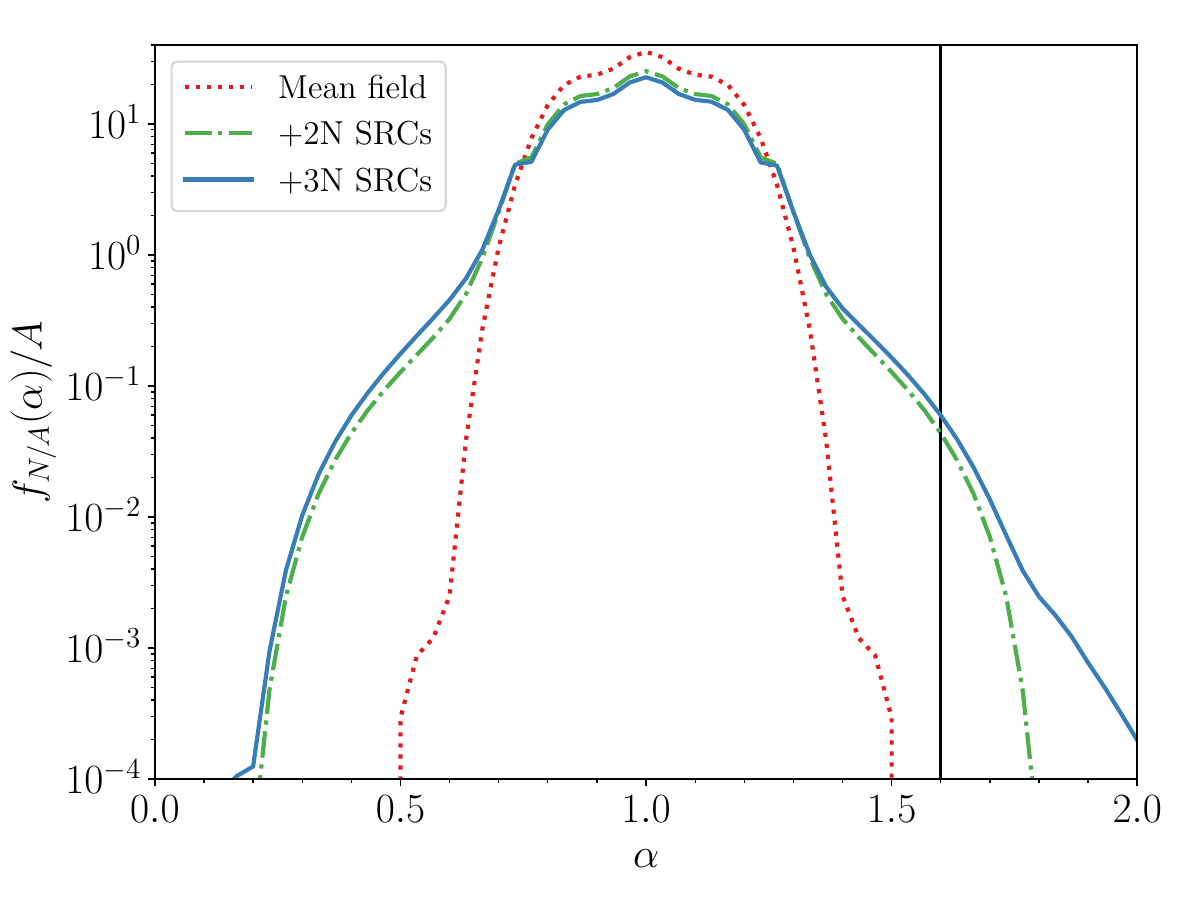}
  \includegraphics[width=0.49\textwidth]{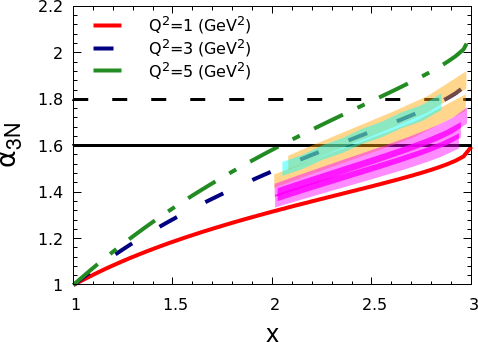}
  \caption{LEFT: Contributions of the mean field, 2N-SRC, and 3N-SRCs, calculated by Ref~\cite{Freese:2014zda}, with a line at $\alpha =1.6$ to indicate the minimum predicted threshold for 3N-SRCs, RIGHT: Magenta band represents data from Ref.~\cite{Ye:3Nsrc}, cyan corresponds to Ref.~\cite{Fomin:2011ng} data, and the yellow bands are forthcoming data from a Hall C experiment~\cite{E12-06-105} at Jefferson Lab. When dealing with experimental kinematics, we use $\alpha_{2N,3N}$ which are approximations to $\alpha$, assuming certain break-up conditions (back-to-back for 2N SRCs, and (B) or (C) in Fig.~\ref{fig:configs} for 3N SRCs.)  See Ref~\cite{Fomin:2017ydn} for details.}
  \label{fig:alpha3Nthres}
\end{figure}

\begin{figure}[htb] 
\centering
  \includegraphics[width=0.98\textwidth,height=0.4
  \textwidth]{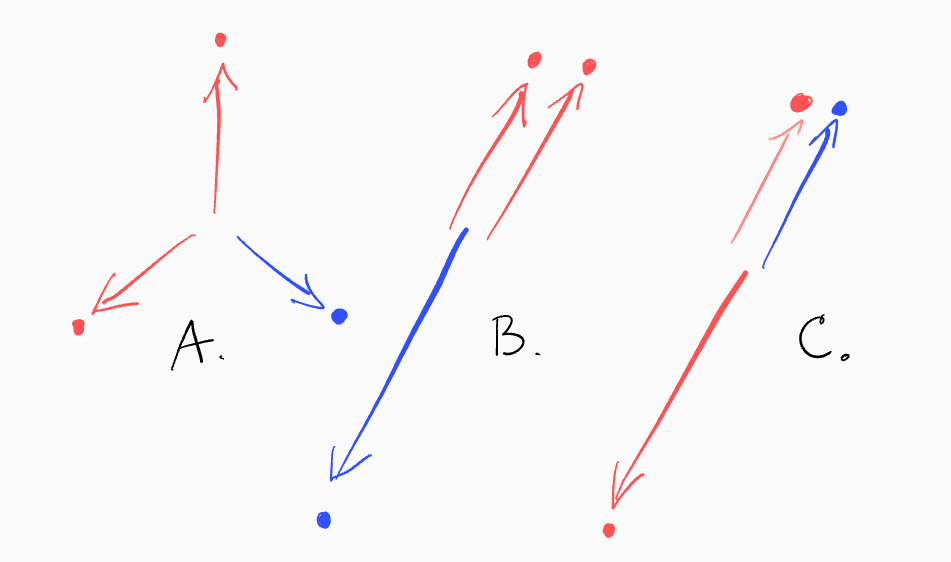}
  \caption{Different possible momentum/isospin configurations for a ppn 3N-SRCs. In (A), the momentum is distributed evenly among all three nucleons, while (B) and (C) have a highly asymmetric sharing of the momentum. When probing the highest-momentum nucleons, (A) will have contributions from two protons and one neutron, while (B) will be dominated by the single neutron and (C) will be dominated by the doubly-occurring proton.}
  \label{fig:configs}
\end{figure}

The presence of 2N-SRCs was established by observing universal two-body behavior in the inclusive cross section ratio for kinematics sensitive to high-momentum nucleons~\cite{Frankfurt:1993sp,CLAS:2003eih,Fomin:2011ng}. Analogous scaling searches have been performed for~\cite{Fomin:2011ng,Ye:3Nsrc,Ye:2018jth} 3N-SRC signatures, expected in the form of a second plateau at higher $x$ values.   There are two main challenges: (1) the contribution from 2N-SRCs falls off much more slowly than the mean field contributions, pushing the 3N-SRC dominance region to more extreme experimental conditions.  Fig.~\ref{fig:alpha3Nthres} (LHS) shows the calculations of mean field, 2N and 3N contributions as a function of $\alpha$ (light-cone momentum fraction of 3N-SRCs carried by the correlated nucleon $i$).  While the 2N contribution begins to noticeably drop off at $\alpha\sim 1.5$, the 3N contribution doesn't become dominant until $\alpha\sim 1.8$ (with $1.6 < \alpha < 1.8$ being a murky region of possibility). The experimental issue is that the cross sections are quite low and run times very long. 
Challenge (2) arises due to the structure of the 3N-SRC system: momentum sharing among the three nucleons can be symmetric or highly-asymmetric, as illustrated in Fig.~\ref{fig:configs}. While we can calculate the minimum momentum of the struck nucleon as a function of $x$ and $Q^2$ for heavy nuclei, this value is dependent on the momentum distribution of the spectators in the 3N-SRC, so a more realistic estimate of the translation between ($x$,$Q^2$) and a minimum momentum for 3N-SRCs can be different for the symmetric and asymmetric momentum configurations.

The most recent guidance for where to focus experimental 3N-SRC searches comes from the authors of Refs~\cite{Sargsian:2019joj,Day2023}. 
 The RHS plot of Fig.~\ref{fig:alpha3Nthres} shows the required experimental conditions to obtain data to see a meaningful second plateau in the cross section ratios.  The semi-transparent bands correspond to existing and forthcoming data from Jefferson Lab experiments in Halls A and C.  If the onset of 3N dominance shown in this figure ($\alpha>1.8$) is correct, it is understandable that no second plateau at x$>$2 has been observed, but recent~\cite{E12-06-105} (and potentially future~\cite{LOI12-21-001}) measurements are to be carried out at kinematics that might be sufficient for an observation.

To expand on challenge (2), this complication arises due to the fact that several isospin configurations are possible for 3N-SRCs. By contract, for 2N-SRCs, where the total momentum is low, the two high-momentum nucleons typically come out of the nucleus back-to-back, making the geometry rather straight-forward.  With an additional third nucleon sharing momentum, there are multiple possible configurations, each of which could look different experimentally.  Even if 3N-SRCs are dominated by ppn and pnn contributions, scattering from isoscalar nuclei should have identical contributions from these configurations (plus possible configurations from ppp- or nnn-SRCs). In the ratio to $^3$He, where only the ppn configuration is possible, the ratio will be sensitive to differences in the number of protons and neutrons at the largest momenta in 3N-SRCs. In a symmetric configuration (left panel of Fig.~\ref{fig:configs}), scattering from the highest momentum nucleons in $^3$He will involve both protons and the neutron.  In a highly asymmetric configuration, (middle and right panels of Fig.~\ref{fig:configs}), the high-momentum part of the distribution could be dominated by the singly-occurring neutron, the double-occurring proton, or instead be a combination of the two. Thus, the question of whether the highest-momentum nucleons are protons, neutrons, or a mix of both will modify the A/$^3$He ratio. If the relative contributions of protons and neutrons varies as a function of momentum, this could also distort the $x$ dependence of the ratio. 

The result of these complications is a situation where it is hard to make reliable predictions for the kinematic ``onset point''. Searches for 3N-SRCs had to take A/$^3$He cross sections at whatever $Q^2$ values were accessible and look for a clear plateau in $x$ for $x$ well above 2, to suppress the 2N-SRCs. As discussed in the next section, existing data do not show indications of 3N-SRCs, but are limited to lower $Q^2$ values or extremely low precision. Therefore, it is critical to extend the $Q^2$ range of such measurements as far as possible, and to find ways to alleviate the issue of corrections associated with different isospin structure in in $^3$He when taking A/$^3$He ratios. One such option, discussed below, is examining ratios of isoscalar nuclei to $^4$He so that we are dominantly seeing equal contributions of ppn- and pnn-SRCs. Additionally, an interesting consequence of the highly asymmetric configurations (middle and right panels of Fig.~\ref{fig:configs} is that comparisons of $^3$H and $^3$He can provide direct information on the momentum and isospin structure of 3N-SRCs. In this case, the highly-asymmetric configurations combined with the single isospin configuration in these A=3 nuclei yield a very different prediction for the cross section ratio if the singly- or doubly-occurring nuclei dominate.

\section{What we learned from previous 3N-SRC searches}
The first observation of a 3N-SRC plateau in A/$^3$He ratios from CLAS data was published in 2006~\cite{egiyan06}. Unlike the 2N-SRC data, these were taken at a single kinematic of $Q^2 \approx 1.6$~GeV$^2$ and the plateau in $x$ appeared to start at $x\sim$2.2.  Both of these features of the data were surprising. First, the 2N-SRC onset was known to begin at $Q^2=$1.4~GeV$^2$, implying a significantly higher value for 3N-SRC observation. Second, the dominance of 3N-SRCs was expected to start higher in $x$, due to a slower fall-off of the 2N-SRC contribution and its center-of-mass momentum.

\begin{figure}[htb!]
   \centering
   \includegraphics[width=0.495\textwidth]{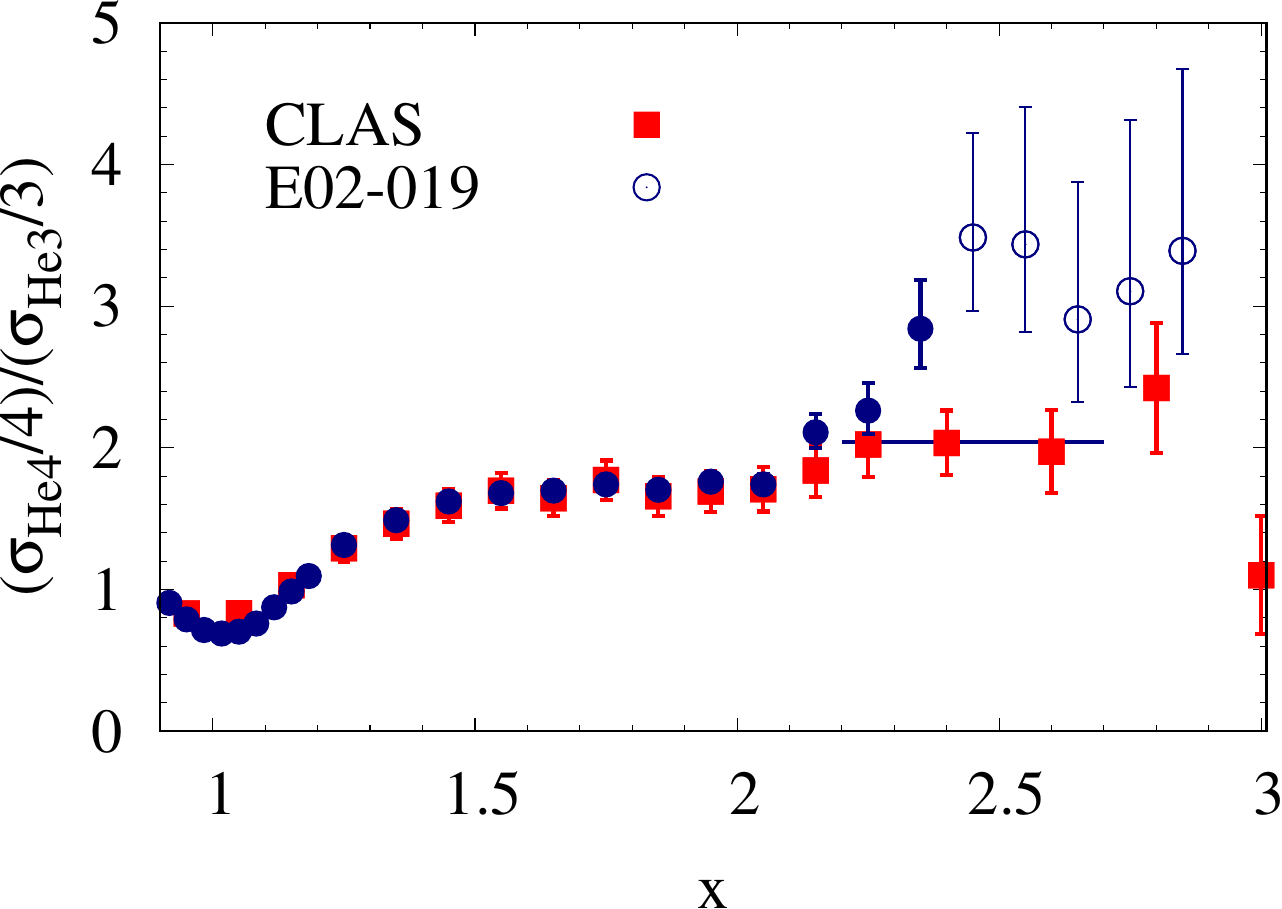} 
  \includegraphics[width=0.475\textwidth]{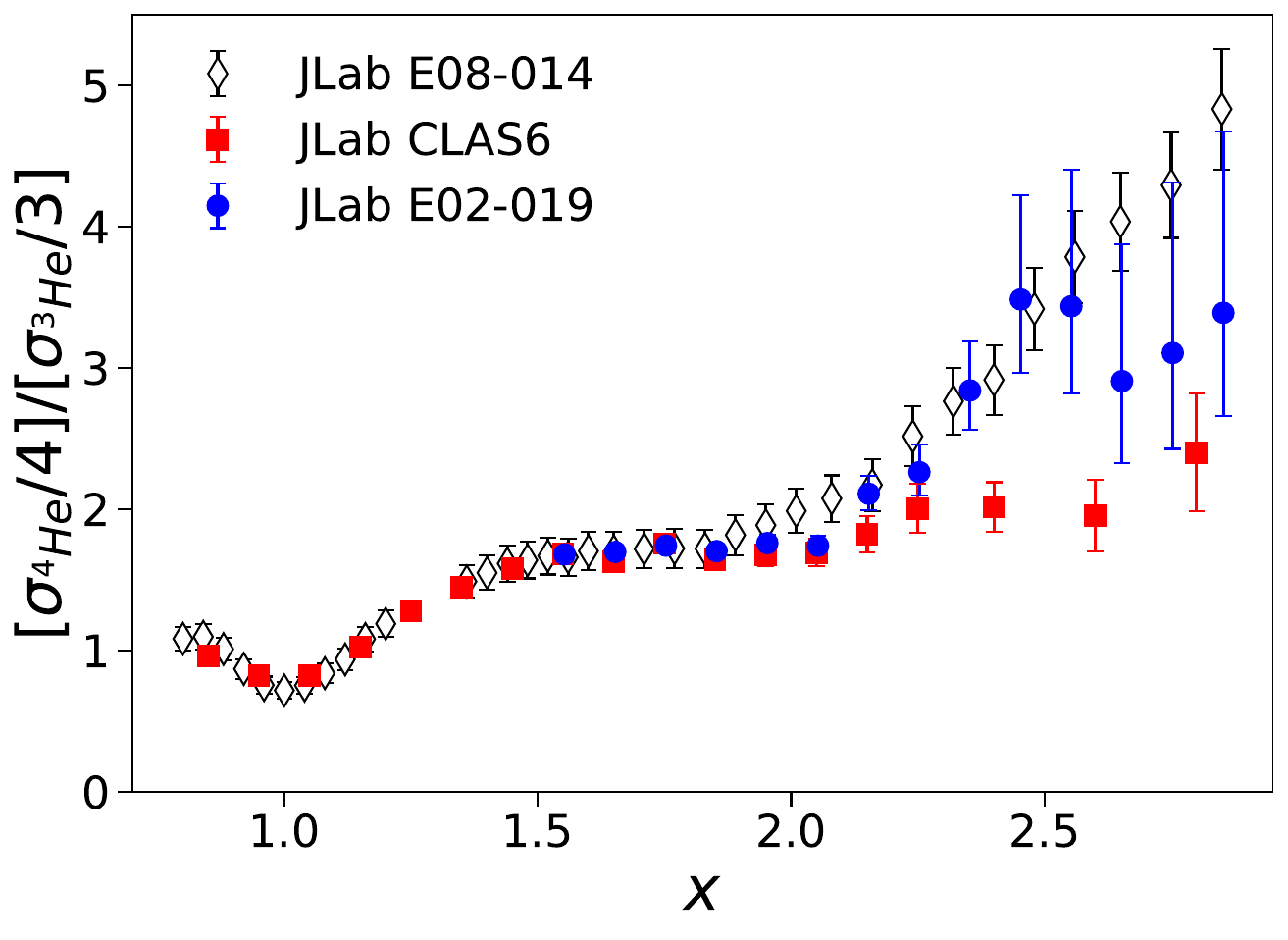} 
   \caption{\textbf{Left:} A/$^3$He cross section ratios from the CLAS measurement ($Q^2 \approx 1.6$~GeV$^2$) and the Hall C E02-019 experiment ($Q^2 \approx 2.9$~GeV$^2$ at $x=2.5$. \textbf{Right:} A/$^3$He ratios for the CLAS and E02-019 experiments, and for the Hall A E08-014 measurement which took data at kinematics approximating the CLAS measurement ($Q^2 \approx 1.6$~GeV$^2$).}
   \label{fig:3n_status}
\end{figure}

An analysis of Hall C 3N-SRC data (E02-019)~\cite{Fomin:2011ng} taken at higher values of $Q^2$ was in progress at the time of the CLAS publication.  These Hall C data were taken at $Q^2 \approx 2.9$~GeV$^2$, almost double that of the CLAS data, implying that the second plateau should be present as well, and its onset should be observed at lower $x$ (an important signature of 2N-SRCs).  However, the statistics on the $^3$He nuclear target did not allow for strong conclusions to be drawn. The experiment utilized a short $^3$He target, leading to a large aluminum endcap subtraction in the kinematic limit.  This, combined with a faster than expected fall-off of the $^3$He cross section at large $x$ yield lowered statistics for the $^3$He data, severely limiting the impact of the A/$^3$He ratios. Despite this fact, the higher-$Q^2$ data were in tension with the previous CLAS 3N-SRC plateau observations~\cite{egiyan06}, as shown in the left panel of Figure~\ref{fig:3n_status}.

With the situation somewhat muddled, the next JLab experiment (E08-014)~\cite{Ye:2018jth} aimed their search at the $x>2$ region with longer cryogenic targets (for higher statistics). Their kinematics were selected with some overlap to those of the CLAS measurement~\cite{egiyan06}, so that they could confirm the initial observation with a high resolution spectrometer as well as to map out the $Q^2$ dependence. Instead, these new data at $Q^2 \approx 1.6$~GeV$^2$ were consistent with the Hall C data at 2.9~GeV$^2$, and disagreed with the CLAS data~\cite{egiyan06} as shown in the right panel of Fig.~\ref{fig:3n_status}. The apparent 3N-SRC plateau was later demonstrated to be an artifact of bin migration associated with the $\sim$1\% momentum resolution in CLAS~\cite{Higinbotham:2014xna}, leaving the question of 3N-SRC existence still open.

The model of Refs.~\cite{Fomin:2017ydn,Sargsian:2019joj,Day2023} also makes a prediction for the probability of finding a nucleon in a 3N-SRC, $a_3(A)$, relative to 2N-SRCs, $a_2(A)$. Because 3N-SRCs are believed to primarily arise as a result of two hard NN interactions, the authors also find that $a_3(A) \propto a_2(A)^2$, assuming that 3N-SRCs are predominantly in ppn or nnp configurations. The authors of Ref~\cite{Sargsian:2019joj,Day2023} go on to test this hypothesis by assuming scaling in this ($\alpha _{3N}$=1.6~-~1.8) region for the E02-019 data and verifying that $a_3 \propto a_2^2$ for these data. This offers additional support that the E02-019 data were consistent with a 3N-SRC scaling plateau, but the result is again limited by the poor statistics of the $^3$He data.  This will be addressed in the forthcoming data from JLab E12-06-105~\cite{E12-06-105}. 

To summarize the status of inclusive 3N-SRC searches: at present, we have a measurement at $Q^2\approx1.6$~GeV$^2$ with moderate statistics that does not show a plateau and data at $Q^2\approx2.9$~GeV$^2$ which are \textit{consistent with a plateau}, but whose precision does not allow for a conclusive statement. 
%
\subsection{XEM2 Experiment}
In this landscape, JLab E12-06-105~\cite{E12-06-105}, a.k.a.``XEM2'', was recently completed in Hall C.  There were several experimental goals, including a 3N-SRC plateau observation.  The dedicated data were taken at $\alpha>1.6$, as guided by the predictions of Refs~\cite{Sargsian:2019joj,Day2023} .   The right panel of Fig.~\ref{fig:alpha3Nthres} shows $\alpha _{3N}$ for different values of $Q^2$ as a function of $x$ to show the relationship between the two variables and where the previous experiments took data in comparison to XEM2. There's overlap between Hall C data from 6 GeV~\cite{Fomin:2011ng} and the data from XEM2, but great care was taken to monitor the statistics during the recent running and long thick high-density helium targets were employed.  This should allow XEM2 to extract highest precision cross section ratios in the 3N-SRC kinematic regime.

XEM2 collected precision data for many nuclear targets~\cite{E12-06-105UPD} (3 $ < $ A $ < $ 232) at 8$^\circ$, corresponding to $Q^2$=1.95~GeV$^2$ at $x=1$ for studies of nuclear and isospin dependence of 2N-SRCs.  Data for 3N-SRC ratios was taken for a subset of targets ($^3$He, $^4$He, $^{12}$C, $^9$Be, and $^{40}$Ca) at 8.5$^\circ$ as well as 10$^\circ$, corresponding to $Q^2$=2.3~GeV$^2$ and $Q^2$=3.2~GeV$^2$ at $x=$2.5, respectively. This is a significant increase in $Q^2$ compared to the Hall A E08-104 measurement~\cite{Ye:3Nsrc}, with improved statistics over both previous measurements.  This will provide a first look at the region where models suggest 3N-SRCs may dominate the scattering cross section. Data at two kinematic settings will also allow for the examination of the $Q^2$ dependence of the 3N-SRC plateau, should it be observed.  Finally, while examining $^{12}$C/$^4$He ratios will yield a smaller signal in $^{12}$C/$^3$He, comparing isoscalar nuclei avoids bias associated with the comparison to $^3$He, which has only ppn configurations.

\section{Isospin studies using $^3$H and $^3$He}\label{sec:tritium}

Comparing measurements of $^3$H and $^3$He provides the cleanest way to study the structure of ppn and pnn configurations. As illustrated in Fig.~\ref{fig:configs}, there are multiple configurations that, in the limit of large nucleon momentum, yield different contributions from protons and neutrons. In the 3N-SRC dominance region, configuration (B) is dominated by the singly-occurring nucleon, so the $^3$H/$^3$He cross section ratio will be close to $\sigma_p/\sigma_n$, configuration (C) is dominated by the doubly-occuring nucleon and will yield roughly $\sigma_n/\sigma_p$, and configuration (A) will be approximately $(\sigma_p + 2\sigma_n)(2 \sigma_p + \sigma_n)$. If we take $\sigma_p / \sigma_n \approx 2.5$ for the kinematics relevant to these studies and assume that the 3N-SRCs are dominated by a single configuration from Fig.~\ref{fig:configs}, we obtain the following predictions:

\begin{eqnarray*}
\sigma_{3H}/\sigma_{3He} & \approx & 0.75~ \mbox{for configuration (A)}, \\
 & \approx & 2.5~ \mbox{for configuration (B)}, \\
 & \approx & 0.4~ \mbox{for configuration (C)}.
\end{eqnarray*}

JLab experiment E12-11-112~\cite{Li:2022fhh} measured $x>2$ $^3$He/$^3$H cross section ratios but its kinematic coverage didn't reach the $\alpha_{3N}=1.6$ threshold and is therefore insufficient to isolate 3N-SRCs.
However, as 3N-SRCs begin to contribute, we would expect the ratio to begin to rise or fall towards the limit obtained for 3N-SRC dominance, providing useful information even in the absence of scaling in the A/$^3$He ratios.

Note that E12-11-112 observed the onset of 2N-SRC scaling behavior at significantly lower $x$ and $Q^2$ than previous measurements. This is in part because of the smaller Fermi momentum for these light nuclei, and in part because scaling-violating effects at low $Q^2$, e.g. meson-exchange contributions and final-state interactions, are expected to be nearly identical in $^3$H and $^3$He. This suggests that the data at $x>2$, even at these $Q^2$ values, may yield a ratio consistent with that expected for 3N-SRC dominance at lower $Q^2$ than for similar studies of A/$^2$H or A/$^3$He ratios.

A similar $^3$H/$^3$He measurement at higher $Q^2$ values, where contributions from 3N-SRCs are dominant, is possible with the high momentum spectrometers at JLab Hall C~\cite{LOI12-21-001}. Based on the XEM2 experiment result, such a direct comparison of $^3$H/$^3$He cross section at $\alpha_{3N}>1.6$ would either allow searching for an $x>2$ plateau with an isoscalar $A=3$ configuration, or provide direct access to the isospin structure of 3N-SRCs.

\section{Perspective on future searches for 3N-SRCs}

This discussion focused on the identification of 3N-SRC dominance in inclusive scattering and isospin studies relying on comparisons of mirror nuclei, $^3$He and $^3$H.  Recently, exclusive measurements (two- and three-nucleon knockout) were performed in CLAS~\cite{E12-17-006A}  as a complementary approach to study the isospin structure of the 3N-SRCs, as was done for 2N-SRCs~\cite{Subedi:2008zz,LabHallA:2014wqo,CLAS:2018xvc,JeffersonLabHallA:2007lly} . 

The recent XEM2 data offers the highest chance of seeing a 3N-SRC plateau in inclusive cross section ratios given its $Q^2$ reach and statistical precision.  While this will be a milestone, it will not be the last step in our 3N-SRC studies.  

In order to provide conclusive evidence of 3N-SRCs, several observations are necessary:  (1) establish $Q^2$ independence required by the SRC picture by observing scaling at more than one scattering angle, (2) obtain direct cross section ratios to $^3$He for theory comparisons, and (3) perform clear tests of the isospin dependence of 3N-SRCs with $^3$He and $^3$H. 

A new experimental Letter of Intent (LOI)~\cite{LOI12-21-001} proposes to do all of the above-mentioned tests.  The first point will require data at an additional (probably higher) $Q^2$ value, wheres the others require longer data taking periods on $^3$He and $^3$H targets.  

In conclusion, recently completed and upcoming inclusive experiments at JLab promise to help elucidate our understanding of short-range nuclear structure in nuclei and extend studies of Short-Range Correlations in nuclei beyond 2N configuration.

\begin{acknowledgements}
This work was supported by U.S. Department of Energy, Office of Science, Office of Nuclear Physics, under contract numbers DE-SC0013615 and DE-AC02-05CH11231.

\end{acknowledgements}


\bibliographystyle{spphys}       
\bibliography{3NSRC_EPJA}   

\end{document}